\documentclass[sigconf,10pt]{acmart}
\usepackage{amsmath,amsfonts}
\usepackage{algorithmic}
\usepackage{algorithm}
\usepackage{array}

\usepackage[caption=false, font=normalsize, labelfont=sf, textfont=sf]{subfig}

\usepackage{textcomp}
\usepackage{stfloats}
\usepackage{url}
\usepackage{verbatim}
\usepackage{graphicx}
\usepackage{bm}
\usepackage{xcolor}
\usepackage[switch]{lineno}
\usepackage{rotating}
\usepackage{enumitem}
\usepackage{listings}
\setlist{nolistsep}

\lstdefinestyle{bashstyle}{
    language=bash,
    basicstyle=\ttfamily\footnotesize,
    backgroundcolor=\color{gray!10},
    frame=single,
    rulecolor=\color{black},
    keywordstyle=\color{blue},
    commentstyle=\color{gray},
    stringstyle=\color{green!50!black},
    showstringspaces=false,
    captionpos=b
}

\begin{document}

\title{How to Increase Energy Efficiency with a Single Linux Command\\ {\small \textbf{Rutgers Department of Computer Science Technical Report DCS-TR-760}}}

\author{Alborz Jelvani}
\affiliation{%
  \institution{Rutgers University}
  \city{Piscataway}
  \country{USA}
}
\email{alborz.jelvani@rutgers.edu}

\author{Richard P Martin}
\affiliation{%
  \institution{Rutgers University}
  \city{Piscataway}
  \country{USA}
}
\email{rmartin@scarletmail.rutgers.edu}

\author{Santosh Nagarakatte}
\orcid{0000-0002-5048-8548}
\affiliation{%
  \institution{Rutgers University}
  \city{Piscataway}
  \country{USA}
}
\email{santosh.nagarakatte@cs.rutgers.edu}

\begin{abstract}
Processors with dynamic power management provide a variety of settings
to control energy efficiency. However, tuning these settings does not
achieve optimal energy savings.
We highlight how existing power capping mechanisms can address these
limitations without requiring any changes to current power
governors. We validate this approach using system measurements across
a month-long data acquisition campaign from SPEC CPU 2017 benchmarks
on a server-class system equipped with dual Intel Xeon Scalable
processors.

Our results indicate that setting a simple power cap can
improve energy efficiency by up to 25\% over traditional energy-saving
system configurations with little performance loss, as most default
settings focus on thermal regulation and performance rather than
compute efficiency. Power capping is very accessible compared to other
approaches, as it can be implemented with a single Linux command. Our
results point to programmers and administrators using power caps as a
primary mechanism to maintain significant energy efficiency while
retaining acceptable performance, as opposed to deploying complex DVFS
algorithms.
\end{abstract}

\renewcommand\footnotetextcopyrightpermission[1]{}

\settopmatter{printacmref=false}
\maketitle
\pagestyle{plain}

\section{Introduction}
It is projected that by 2030 data centers could account for over 9\%
of electricity consumption in the United States \cite{epri}. To reduce
both the costs and the environmental impact of data centers,
optimizing data center energy efficiency is paramount to ensure a
sustainable future. Recent estimates suggest that approximately 20\%
of Microsoft Azure datacenter operational carbon emissions are from
CPUs alone \cite{wang2024designing}.

Given the significant share of energy consumed by CPUs, a troubling
truth is that current CPU power management techniques fall short of
optimizing for energy efficiency. It is well known that power control
algorithms (governors) available within Linux and even those
integrated within processor hardware each fail to maximize energy
efficiency for various workloads. Numerous works
(\cite{spiliopoulos2011green,hebbar2022pmu, huang2024powersave}) have
proposed using performance metrics available in hardware (such as
stalled CPU cycles) to create custom power governors that achieve better
energy efficiency than power governors in Linux and hardware. However,
the prior methods have some of the following limitations. First, the
requirement to run custom code in kernel mode to control CPU frequency
limits accessibility and the custom code may not be rigorously tested
as the official mainline Linux kernel code. Second, controlling CPU
frequency through software adds the overhead of recurring OS
intervention into hardware. On the other hand, hardware-based power
governors can operate without OS intervention and provide quicker
response times to frequency scaling demands.

To overcome the limitations of existing and proposed dynamic voltage
and frequency scaling (DVFS) governors, we instead propose using power
capping mechanisms as a means of increasing system energy efficiency
for CPU workloads. More importantly, we want to emphasize that Linux
and processor hardware already possess power capping mechanisms
capable of increasing energy efficiency.

Our key insight is that most default power cap settings do not
optimize for compute efficiency. The proposed approach of enforcing a
power cap for energy efficiency optimization is transparent to
programmers and administrators, accessible, and has stable performance
vs. efficiency trade-off properties that can be embodied in simple
rules-of-thumb. For example, a simple rule of thumb could be
\textit{``set the power cap to 80\% of the processors thermal design
  power (TDP), unless users complain the system is too slow''}.

This paper empirically shows that programmers and administrators can
use power caps as a primary mechanism for increasing energy efficiency
without compromising performance, achieving energy efficiency
improvements of up to 25\% for some workloads. We employ a test rig
consisting of a server-class system with power supply power
measurements and collect system statistics across various system
configurations for SPEC CPU 2017 benchmarks. We present aggregated
results from a month-long data acquisition campaign. We analyze and
attribute the observed energy efficiency gains to a decrease in
stalled CPU cycles and the possibility of the energy/frequency
convexity rule \cite{de2014energy}.

\iffalse
\begin{itemize}
    \item We empirically show that programmers and administrators can
      use power caps as a primary mechanism for increasing energy
      efficiency without compromising performance, achieving energy
      efficiency improvements of up to 25\% for some workloads.
    \item We employ a test rig consisting of a server-class system
      with power supply power measurements and collect system
      statistics across various system configurations for SPEC CPU
      2017 benchmarks. We present aggregated results from a month-long
      data acquisition campaign.
    \item We analyze and attribute the observed energy efficiency
      gains to a decrease in stalled CPU cycles and the possibility of
      the energy/frequency convexity rule \cite{de2014energy}.
\end{itemize}
\fi

\section{Background}

\subsection{Where does compute power go?}
The power consumption of a CPU is composed of \textit{dynamic} and
\textit{static} power. We can represent the total power consumption of
a CPU as

\begin{equation}\label{eq:1}
    P_{cpu} = P_{dynamic} + P_{static}
\end{equation}
\begin{equation}\label{eq:2}
    =  \alpha CV^2f + V(ke^\beta)
\end{equation}

where $\alpha$ represents the activity factor and is circuit specific,
$C$ represents the load capacitance, and $V$ is the supply
voltage. The additional term, $ke^\beta$, models leakage
currents\footnote{The dominating sources of leakage current include
sub-threshold leakage, gate-oxide tunneling leakage, and
band-to-band-tunneling leakage \cite{agarwal2005leakage}.} which
depends on factors such as temperature and transistor feature sizes
\cite{haj2018power}. From this simplified model, lowering the clock
frequency ($f$) grants dynamic power savings, while static power usage
remains constant.

\subsection{Exploiting Dynamic Power}
As shown in (\ref{eq:2}), $P_{dynamic}$ of the processor is
proportional to the square of the supply voltage ($V$), thus
decreasing $V$ (voltage scaling) grants considerable power
savings. With CMOS transistors, scaling $V$ affects signal propagation
delay and thus higher frequencies require higher supply voltages to
enable faster transistor switching.

A common method used by processors to exploit dynamic power is dynamic
voltage and frequency scaling (DVFS). By controlling voltage-frequency
pairs supplied to each CPU, DVFS can reduce overall system power
consumption at the cost of performance.

To realize DVFS, processors provide subsystems for placing cores into
various power states as defined in the ACPI specification
\cite{acpi}. The ACPI power states used for DVFS are known as
performance states (P-states), which are controlled by \textit{scaling
  drivers} that communicate with the processor's power management unit
(PMU). Each scaling driver provides \textit{scaling governors} which
use metrics such as core utilization to select P-states. Linux
provides scaling drivers such as \texttt{acpi\_cpufreq} and
\texttt{intel\_cpufreq}. Each scaling driver provides governors
integrated into the kernel that monitor CPU utilization and adjust
P-states depending on current throughput demands from running
processes.

Starting with Intel Skylake architecture, scaling governors were
integrated into the hardware PMU to allow rapid P-state transitions
without OS intervention \cite{doweck2017inside}, debuting the term
\textit{autonomous frequency scaling}. PMU-based scaling drivers (such
as \texttt{intel\_pstate}) are now the default governors for Linux
running on Intel processors and provide two pseudo-governors:
\texttt{performance} and \texttt{powersave}. The former hints to the
PMU to operate at the highest possible frequency while the latter
hints to the PMU to select the lowest frequency proportional to the
current workload \cite{pstate_kernel}.

\subsection{Power Capping}
While DVFS can scale CPU voltage and frequency to manage power
consumption, controlling frequency directly affects the performance of
a CPU. An alternate abstraction for power management of a CPU is
through power capping. Power capping allows a user to assign a power
limit to the processor; the power capping framework enforces this
power limit through methods such as DVFS. One advantage of this
approach is that it grants users greater control over the PMU's
frequency scaling algorithm compared to the pseudo-governors offered
by manufacturers like Intel.

One such power capping framework is Intel's Running Average Power
Limit (RAPL), which was introduced with the Intel Sandy Bridge
architecture \cite{rotem2012power}. RAPL provides power capping
mechanisms for various subsystems across the processor (known as
\textit{power zones}) which are controllable through model-specific
registers (MSRs). The RAPL framework allows monitoring and power
capping of each power zone (such as core, uncore, and DRAM). To use
RAPL, a user specifies a wattage power limit and time window for a
power zone. RAPL then ensures the average power usage of the power
zone does not exceed the power limit within the time window. The Linux
kernel provides an interface to RAPL's MSR registers through the
\texttt{sysfs} filesystem. This effectively allows user space to
modify the power limit of the processor and also monitor energy usage
for various power zones.

The default RAPL configuration for the system we utilize in the
experiments is presented in Listing \ref{fig:rapl_config}. As shown,
this system has 2 power zones (\texttt{Zone 0} and \texttt{Zone 1}),
one for each socket. Each power zone has \texttt{long\_term}
constraint with a default power cap of 150 watts and a time window of
$999424 \mu s$. The \texttt{long\_term} power cap is set by default to
the processors TDP. The time window defines the duration over which
RAPL averages power consumption to ensure that it remains within the
specified power cap. Each power zone also has a \texttt{ short\_term}
constraint with a default power cap of 180 watts and a time window of
$1952 \mu s$. This second power cap is used when Intel Turbo Boost is
active.

RAPL was originally created to manage and cap server power and thermal
loads in data centers, enabling more predictable resource utilization
\cite{david2010rapl}. Since its inception, RAPL has been used for
estimating CPU and memory power consumption (\cite{khan2018rapl,
  kambadur2014experimental}) and managing power consumption in data
centers (\cite{wu2016dynamo, patki2013exploring,
  kumbhare2021prediction, lo2014towards, zhang2021flex}).

In this work, we utilize RAPL to vary the power limit of the entire
processor and measure energy efficiency gains for each system
configuration at varying core counts. Our results indicate that using
RAPL to decrease the power limit of the processor over the default
values (effectively this is the TDP of the processor) can
significantly improve energy efficiency, indicating that the lack of
workload aware power management in current PMU power governors can be
corrected through existing hardware and software mechanisms.

Although this method was partially explored by
Kambadur~et~al. \cite{kambadur2014experimental}, we look at the
effects of power capping on energy efficiency in more detail and show
power capping can increase energy efficiency by a much larger margin
than previously known.

\section{Methodology} \label{methodology}

\begin{table}[t]
    \caption{Server Specifications}
    \centering
    \begin{tabular}{|l|l|}
        \hline
        \multicolumn{2}{|c|}{\textbf{System Specifications}} \\
        \hline
        System & Dell PowerEdge R740 \\
        \hline
        Processor & Dual Intel Xeon Gold 6242 \\
        \hline
        Feature Size & 14 nm \\
        \hline  
        Core Count & 32 (64 logical cores)\\
        \hline
        Frequency (GHz) & 1.20 - 3.90 \\
        \hline
        TDP & 150W (per socket) \\
        \hline
        Memory & 384 GiB ECC DDR4 2933 MHz\\
        \hline
        Default RAPL Power Limit & \texttt{long\_term}: 150W \\ & \texttt{short\_term}: 180W \\
        \hline
        Power Supplies & Dual redundant 750W \\
        \hline
        Software & Ubuntu 22.04.4 LTS, Linux 5.15\\ & gcc 11.4.0\\
        \hline
        \multicolumn{2}{|c|}{\textbf{BIOS Configuration}} \\
        \hline
        Turbo Boost & Enabled \\
        \hline
        CPU Power Management & OS DBPM \\
        \hline
        Energy Efficient Policy & Energy Efficient \\
        \hline
        iDRAC Thermal Profile & Performance per Watt \\ &
        Optimized \\
        \hline
        iDRAC Power Cap & Disabled \\
        \hline
    \end{tabular}
    \label{tab:server_specs}
\end{table}

We perform experiments on a Dell Poweredge R740 (detailed specifications are provided in Table \ref{tab:server_specs}) server and present our findings based on data gathered across a month-long data acquisition campaign.

We run benchmarks from the SPEC CPU 2017 Speed Suite and collect system-wide metrics using a telemetry collection script that samples CPU core frequencies and RAPL energy counters at 10hz. We also validate these results by measuring the power consumption of the entire server from the power supply using the Intelligent Platform Management Interface (IPMI) interface available on the server; we take the integral of the power measurements with respect to time to obtain total energy consumption. Each benchmark is run with the SPEC reference input size. For each run of a benchmark we vary the enabled core count on the system by setting the \texttt{sysfs} attribute accessible at \texttt{/sys/devices/system/cpu/online} and run the benchmark with a thread count equal to the core count. We also utilize \texttt{sysfs} to vary the \texttt{short\_term} and \texttt{long\_term} RAPL power limit for each socket’s package domain power zone, ranging from 70W to 180W in 10W increments. For both sockets, we set the \texttt{long\_term} and \texttt{short\_term} constraints to the same power limit. A sample of the relevant Bash commands is shown in Listing \ref{fig:bash_script}.

 To quantify efficiency gains over the current processor mechanisms for tuning energy efficiency, we use the \\ \texttt{intel\_pstate} CPU power scaling driver with the \texttt{powersave} power governor. Additionally, we set the Intel Performance and Energy Bias Hint (EPB) value to $15$ on all cores to maximize energy savings as managed internally by the processor. Furthermore, we set various BIOS settings to optimize for energy efficiency as outlined in Table \ref{tab:server_specs}.

\section{Evaluation}

\begin{figure}[t]
    \centering
  \subfloat[Processor (RAPL) energy efficiency matrix.\label{fig:emat_rapl}]{%
       \includegraphics[width=1\linewidth]{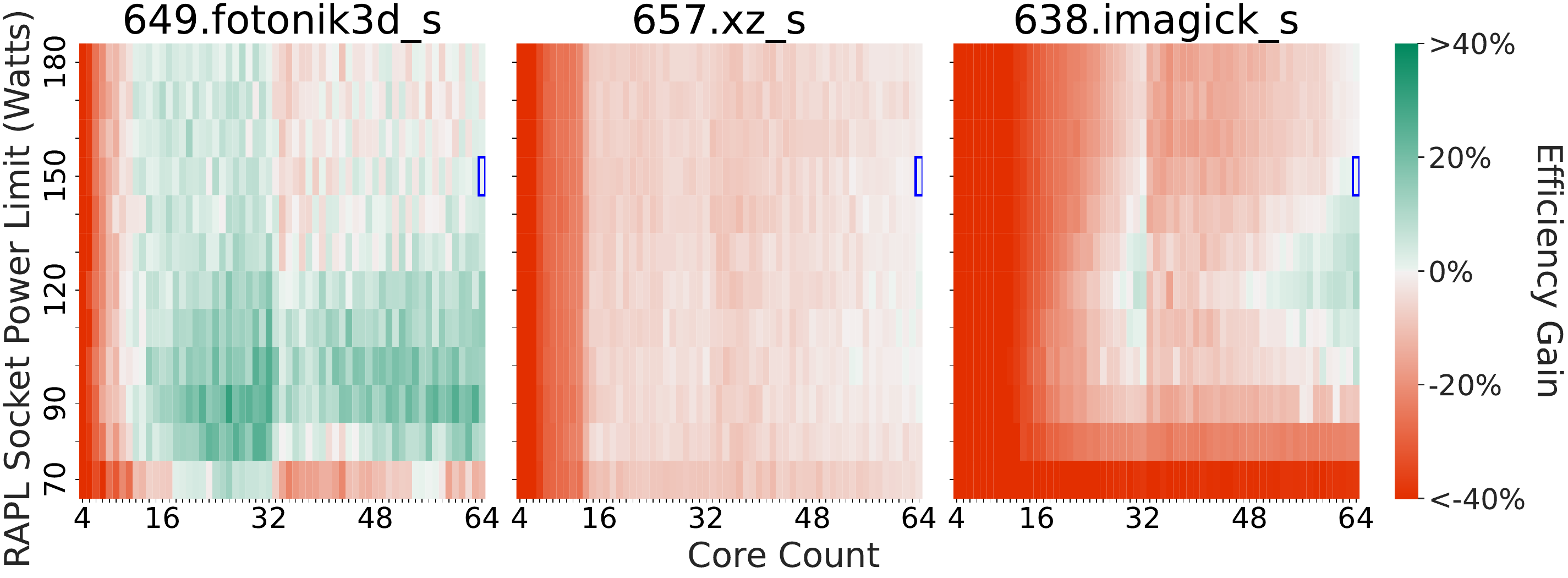}}
    \\
  \subfloat[Entire server (IPMI) energy efficiency matrix.\label{fig:emat_ac}]{%
       \includegraphics[width=1\linewidth]{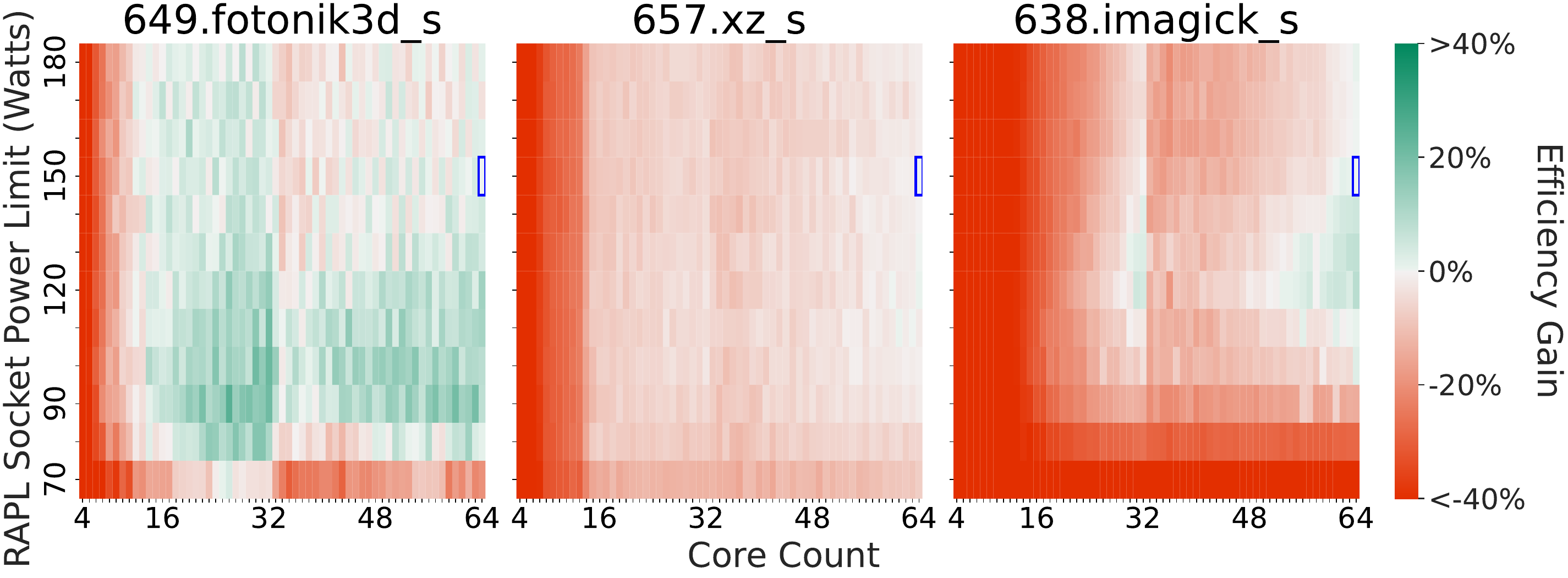}}
    \\
  \subfloat[Performance matrix.\label{fig:pmat}]{%
       \includegraphics[width=1\linewidth]{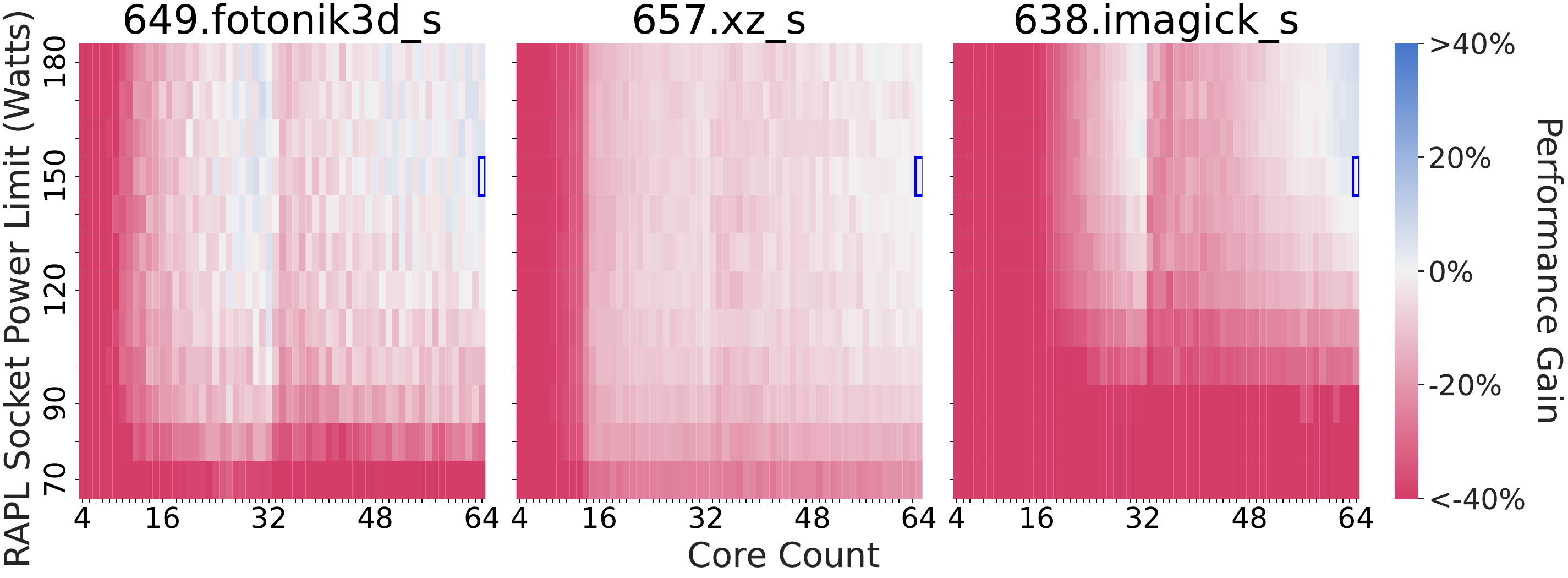}}

  \caption{(a), (b) Efficiency and (c) performance gains for SPEC CPU 2017 benchmarks at various RAPL power limits and core counts normalized against the baseline system configuration in Listing \ref{fig:rapl_config} (blue box).}

  \label{fig:epmat}
\end{figure}

\begin{figure}[h]
    \centering
    \subfloat[Proportion of stalled cycles across RAPL power limits for the top 5 benchmarks from Figure \ref{fig:stallsdiff}.\label{fig:stalls}]{%
       \includegraphics[width=1\linewidth]{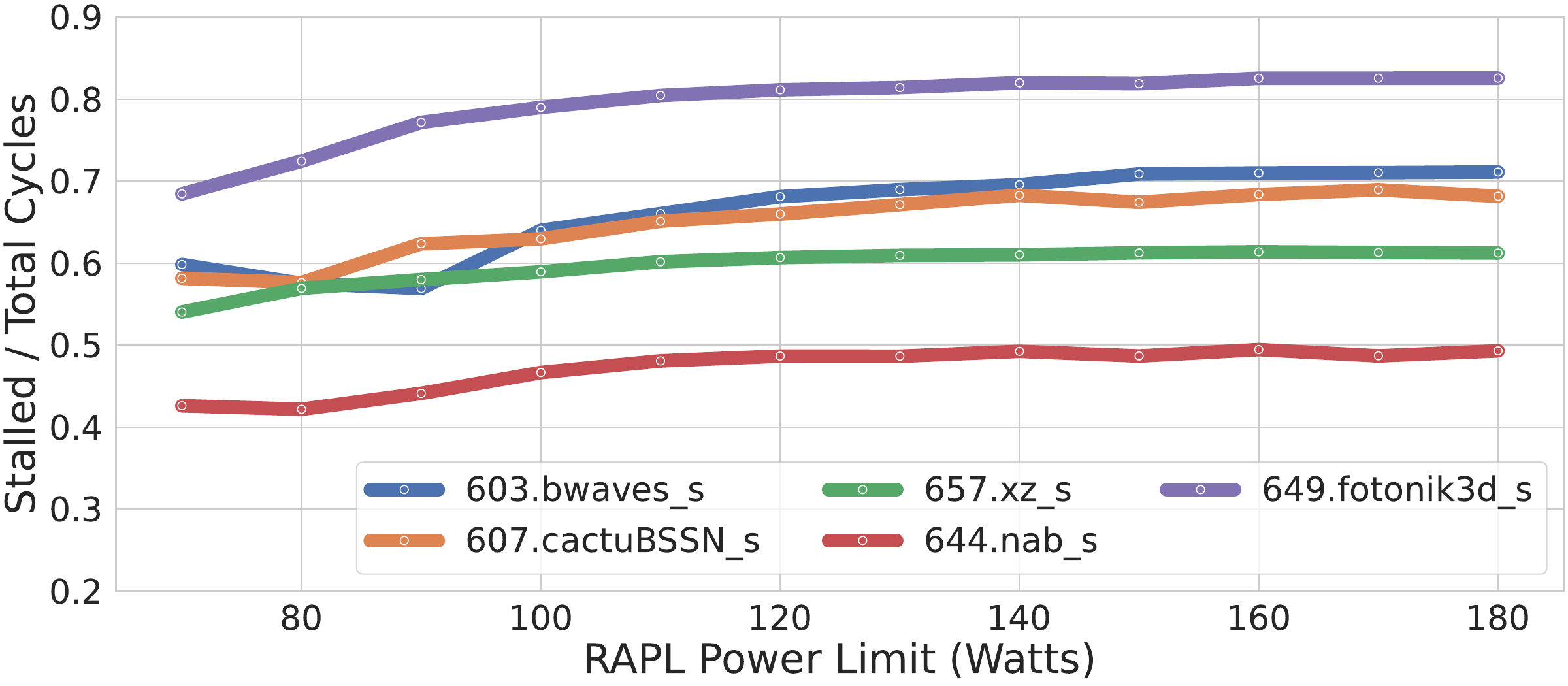}}
    \\
    \subfloat[Range (minimum and maximum) of stalled cycles achievable through power capping. Bottleneck classification of each benchmark is based on the work by Hebbar at al. \cite{hebbar2022pmu}.\label{fig:stallsdiff}]{%
       \includegraphics[width=1\linewidth]{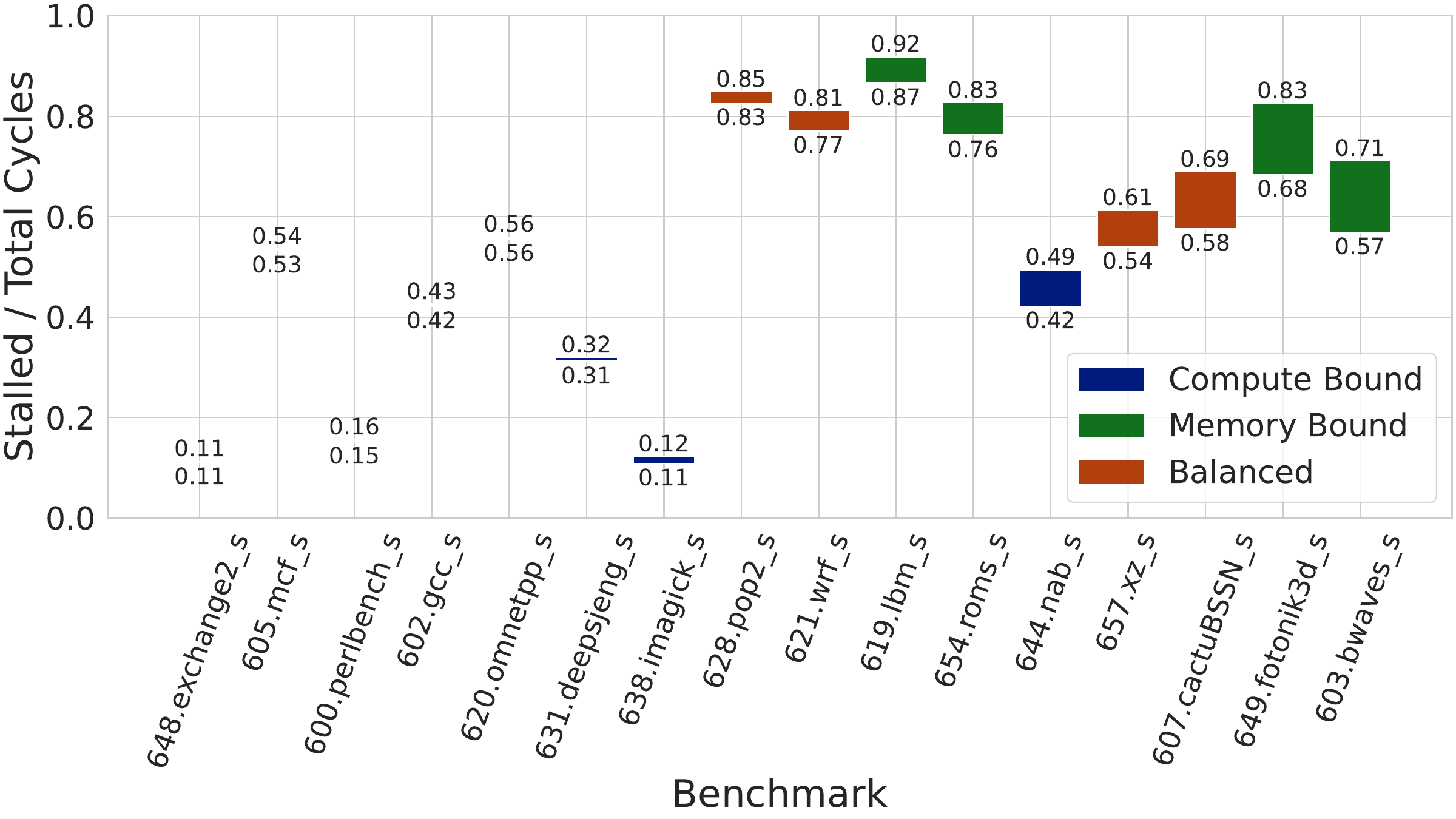}}
    \\
    \caption{Stalled cycles for SPEC CPU 2017 Speed benchmarks ran with 64 cores varied across power limits of 70-180w and averaged over 5 trials.}

\end{figure}

We run benchmarks from the SPEC CPU 2017 Speed suite; for these benchmarks runtime determines absolute performance. The efficiency and performance matrices for \textit{649.fotonik3d\_s}, \textit{657.xz\_s}, and \textit{638.imagick\_s} is shown in Figure \ref{fig:epmat}. Each cell of the efficiency matrices contains normalized energy usage for a single run of the benchmark at a given core count and RAPL power limit while each cell of the performance matrix contains the normalized runtime. We normalize energy efficiency and run times to the default system configuration of 64 enabled cores and a RAPL power limit of 150W (marked with a blue box). The default RAPL configuration used for the reference configuration is presented in Listing \ref{fig:rapl_config}.

Figure \ref{fig:epmat} shows that for the SPEC CPU benchmarks, setting the RAPL limit to 120 watts improves energy efficiency for two applications with very little performance loss. If such a power limit exists for many more applications, it points the way for a rule-thumb that users can apply to RAPL settings where many applications gain significant energy efficiency, on the order of 20-25\% with only a few percentage points loss in run-time latency.

\begin{figure*}[t]
    \centering
   \includegraphics[width=1\textwidth,trim={0 0 0 0},clip]{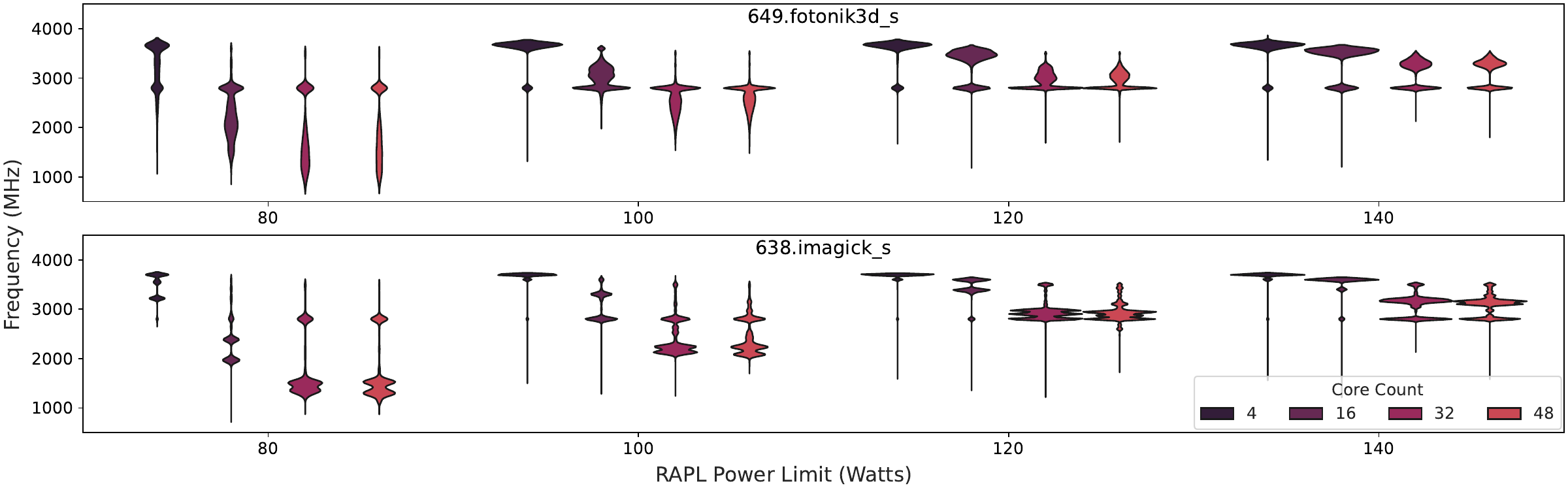}
    \caption{Violin plots illustrating the distribution of core frequencies at various RAPL power limits and core counts for \textit{649.fotonik3d\_s} and \textit{638.imagick\_s}. Each violin contains core frequencies from a trace for a corresponding cell from Figure \ref{fig:epmat}. Increasing core counts saturate the corresponding RAPL power budget faster, resulting in lower frequencies in the violin plots as core counts increase.}
    \label{fig:violin}
\end{figure*}

We also present a snapshot of the core frequencies for select RAPL power limits and enabled core counts in Figure \ref{fig:violin}. Each violin shows the distribution of core frequency readings for a single RAPL power limit and core count configuration.

\subsection{Analysis}

Our results in Figure \ref{fig:epmat} show a clear distinction in energy efficiency gains for various system configurations depending on workload characteristics, such as memory-intensive workloads (\textit{649.fotonik3d\_s}), balanced workloads (\textit{657.xz\_s}), and compute-intensive workloads (\textit{638.imagick\_s}). As shown, we can trade a \textit{bit} of performance for much greater energy efficiency gains through power capping, something the existing power-management interface on the processor is unable to optimize for even with various energy efficiency knobs tuned.

\subsubsection{\textbf{Impact of Power Capping on Energy Efficiency Across Workload Types}}

For memory-intensive workloads such as \textit{649.fotonik3d\_s}, using a low power cap for each socket can result in up to $25\%$ in energy savings, even with all processor power tuning options set to optimize for energy efficiency. \textit{More specifically, we can gain $25\%$ in energy efficiency while trading less than $5\%$ in performance (at a power cap of 90W with 26 cores enabled)}. This result contrasts from the study by Kambadur et al. \cite{kambadur2014experimental} which observed a net energy increase across various power caps on an Intel Sandybridge system.

On the other hand, a balanced workload such as \textit{657.xz\_s} achieves no considerable energy efficiency gain with the same power capping modifications. Interestingly enough, a compute-intensive workload such as \textit{638.imagick\_s} achieves better energy efficiency at low power caps when all cores in each socket are utilized. For example, as shown in Figure \ref{fig:epmat}, we can trade a $7\%$ performance loss for a $9\%$ gain in energy efficiency (at a power cap of 120 watts with 64 cores enabled). For all three workloads, a clear efficiency and performance drop is apparent when the 33rd core is enabled, as this enables the second socket in the system.

\subsubsection{\textbf{Why Lower Power Caps Improve Efficiency in Select Workloads}}

Although faster runtimes \textit{may} lead to improved energy efficiency, this is not always true, as demonstrated by the performance matrix in Figure \ref{fig:pmat}. 
Performance gains are only possible for \textit{649.fotonik3d\_s} and \textit{638.imagick\_s} (the blue shaded regions), and in both instances the gains are less than $10\%$ and appear mostly where expected (at high core counts and power caps).

An explanation for the energy efficiency gains in \textit{649.fotonik3d\_s} can be explained by the memory-intensive nature of the benchmark. Figure \ref{fig:stallsdiff} illustrates how the ratio of stalled to total CPU cycles varies for the SPEC CPU 2017 Speed benchmarks under different RAPL power caps. Memory-intensive benchmarks such as \textit{649.fotonik3d\_s} exhibit a wide range of stalled cycles depending on the power limit utilized. 

Figure \ref{fig:stalls} plots the stalled cycle ratios at various RAPL power limits for the five SPEC benchmarks with the largest ranges, as identified in Figure \ref{fig:stallsdiff}. For all of these benchmarks, the stalled cycle ratio generally increases with higher power caps and converges to a steady-state value. The existence of a steady-state for the stalled cycle ratio can be attributed to the increase in the power budget for each core which impacts the operation of DVFS under a power cap and can be partially observed in Figure \ref{fig:violin}. 

In the violin plot for \textit{649.fotonik3d\_s}, the violins for a RAPL power cap of 80W exhibits wider variability in the recorded core frequencies compared to the violins for a RAPL power cap of 140W, which exhibit core frequencies that tend to the upper envelope of 3.9GHz. \textit{At the lower power caps cores saturate the power budget and RAPL throttles frequencies, which balances the compute bandwidth with the memory bandwidth of the system, decreasing stalled cycles.} At higher power caps the cores do not fully saturate the power budget; any core throttling due to memory bandwidth saturation is controlled purely by the processors PMU governor and not RAPL.

\subsubsection{\textbf{Effects of Frequency Scaling on Energy Efficiency for Compute Intensive Workloads}}

A more challenging result to explore are the gains in energy efficiency we observed for \textit{638.imagick\_s} as shown in Figures \ref{fig:emat_rapl} and \ref{fig:emat_ac}. The energy efficiency gains were obtained at a higher cost of performance. This means that as energy efficiency increased, performance decreases were greater than those of memory-intensive benchmarks. Recall in Figure \ref{fig:epmat}, the most energy optimal configuration for \textit{638.imagick\_s} was at a 120 watt power cap with 64 enabled cores. This traded a $7\%$ performance loss for a $9\%$ gain in energy efficiency. Furthermore, Figure \ref{fig:stallsdiff} shows that the range of the stalled cycle ratio for \textit{638.imagick\_s} is almost unchanged when power limits are varied; thus stalled CPU cycles are not to blame for the decrease in energy efficiency.

A possible explanation for the energy efficiency gain for compute intensive workloads under a power cap may be hidden in Figure \ref{fig:violin}. The violin plot for \textit{638.imagick\_s} indicates that average core frequencies decrease as core counts increase. Additionally, the RAPL power limit determines the rate of decrease for core frequencies as core count increases (this can be attributed to a faster saturation of the power budget for the lower power limit as opposed to the higher power limit). 

The throttling of the core frequencies at the lower power limits and the nearly smooth features of the efficiency matrix in Figure \ref{fig:epmat} for \textit{638.imagick\_s} may be evidence of the energy/frequency convexity rule \cite{de2014energy}, the property that scaling the frequency for fixed workloads exhibits a convex energy consumption curve (i.e., there is an energy-optimal CPU frequency for an arbitrary workload). The convexity rule is partially attributed to the quadratic scaling of power as voltage linearly increases along with the constant static power leakage irrelevant of frequency (Equation \ref{eq:2}).

If the energy/frequency convexity rule is indeed at play here, it is important to note that RAPL can account for static and dynamic power leakage as both the efficiency matrices of the entire server (Figure \ref{fig:emat_ac}) and RAPL (Figure \ref{fig:emat_rapl}) exhibit a similar optimal region for \textit{ 638.imagick\_s} \footnote{This also means other server components (e.g., voltage regulators, power supply, etc) do not disproportionately affect overall system efficiency.}. This indicates that although hardware-based DVFS may be failing to exploit the convexity rule, it has access to the relevant telemetry required to do so.
\section{Conclusion}
Prior studies have shown that existing processor power management
techniques often lack workload-aware power scaling
\cite{hebbar2022pmu, lo2014towards,
  spiliopoulos2011green,imes2019copper}, and in some cases may even
exhibit behavior contrary to their intended design
\cite{huang2024powersave}.
Although developing a custom power governor is a viable approach to
improve processor power management, our results show that significant
energy efficiency gains can be achieved using power capping mechanisms
readily available on modern server processors. These mechanisms are
easily accessible, and setting appropriate power caps could become
standard practice for system administrators with minimal impact on
user experience.

\bibliographystyle{acm-ref}
\bibliography{refs}

%\appendix
%\clearpage
\onecolumn
%\section{Appendix}
\begin{figure}[!htbp]
\centering
    \begin{lstlisting}[style=bashstyle, numbers=left, caption={Bash script snippet for setting the RAPL power limit for both sockets on the Dell PowerEdge R740.}, label={fig:bash_script}]
#!/usr/bin/env bash

# set rapl power limit
# usage ./rapl_set.sh <PLIMIT>
# where <PLIMI> is in watts

microwatts=$(($1 * 1000000))

echo $microwatts | sudo tee /sys/class/powercap/intel-rapl/intel-rapl\:0/constraint_0_power_limit_uw
echo $microwatts | sudo tee /sys/class/powercap/intel-rapl/intel-rapl\:0/constraint_1_power_limit_uw
echo $microwatts | sudo tee /sys/class/powercap/intel-rapl/intel-rapl\:1/constraint_0_power_limit_uw
echo $microwatts | sudo tee /sys/class/powercap/intel-rapl/intel-rapl\:1/constraint_1_power_limit_uw

printf "RAPL limit set to $1 watts\r\n"
    \end{lstlisting}
\end{figure}
\begin{figure}[t]
\centering
    \begin{lstlisting}[style=bashstyle, caption={The default RAPL configuration of the Dell PowerEdge R740.}, label={fig:rapl_config}]
Zone 0
  name: package-0
  enabled: 1
  max_energy_range_uj: 262143328850
  Constraint 0
    name: long_term
    power_limit_uw: 150000000
    time_window_us: 999424
    max_power_uw: 150000000
  Constraint 1
    name: short_term
    power_limit_uw: 180000000
    time_window_us: 1952
    max_power_uw: 376000000
  Subzone 0
    name: dram
    enabled: 0
    max_energy_range_uj: 65712999613
    Constraint 0
      name: long_term
      power_limit_uw: 0
      time_window_us: 976
      max_power_uw: 41250000
Zone 1
  name: package-1
  enabled: 1
  max_energy_range_uj: 262143328850
  Constraint 0
    name: long_term
    power_limit_uw: 150000000
    time_window_us: 999424
    max_power_uw: 150000000
  Constraint 1
    name: short_term
    power_limit_uw: 180000000
    time_window_us: 1952
    max_power_uw: 376000000
  Subzone 0
    name: dram
    enabled: 0
    max_energy_range_uj: 65712999613
    Constraint 0
      name: long_term
      power_limit_uw: 0
      time_window_us: 976
      max_power_uw: 41250000
    \end{lstlisting}
\end{figure}

%\twocolumn

\end{document}